**Comment on "Minimal Basis Iterative Stockholder: Atoms in Molecules for Force-Field Development"**

Thomas A. Manz, Department of Chemical & Materials Engineering, New Mexico State University, Las Cruces, New Mexico, email:tmanz@nmsu.edu

### Abstract

Verstraelen et al. (J. Chem. Theory Comput. 12 (2016) 3894-3912) recently introduced a new method for partitioning the electron density of a material into constituent atoms. Their approach falls within the class of atomic population analysis methods called stockholder charge partitioning methods in which a material electron distribution is divided into overlapping atoms. The Minimal Basis Iterative Stockholder (MBIS) method proposed by Verstraelen et al. composes the pro-atom density as a sum of exponential functions, where the number of exponential functions equals that element's row in the Periodic Table. Specifically, one exponential function is used for H and He, two for Li through Ne, three for Na through Ar, etc. In the MBIS method, the exponential functions parameters defining the pro-atom density are optimized in a self-consistent iterative procedure. Close examination reveals some important anomalies in the article by Verstaelen et al. The purpose of this comment article is to bring these important issues to readers' attention and to start a discussion of them.

### I. Introduction

Quantum chemistry methods such as Density Functional Theory (DFT), coupled cluster theory, configuration interaction methods, etc. can calculate a material's geometry, electron density distribution, and spin distribution. A key question is how to partition a material's electron density distribution into constituent atoms. This question is both challenging and important. It is challenging, because the continuous nature of the electron cloud means there is some flexibility in how to define such a partition. It is important, because many of the derived properties such as net atomic charges are widely used throughout the chemical sciences to understand and computationally model material properties.

There are several major classes of charge assignment methods. Briefly, we could envision four:

(a) those which divide the total electron distribution, $\rho(\vec{r})$, into non-overlapping atomic compartments (e.g., Bader's quantum theory of atoms in molecules (QTAIM)[2]),

(b) those which divide the total electron distribution into overlapping atoms with non-negative atomic electron distributions,

(c) those which divide the total electron distribution into overlapping atoms but without the constraint that the assigned atomic electron distributions should be non-negative, and

(d) those which assign integrated properties (such as net atomic charges) but do not yield corresponding atomic electron density distributions that sum to the total electron density.

Approach (b) is the class of atomic population analysis methods called stockholder charge partitioning methods in which the electron distribution assigned to each atom A in a material can be represented by

$$\rho_A(\vec{r}) = \rho_A^0(\vec{r}) \rho(\vec{r}) / \rho^0(\vec{r}) \qquad (1)$$

where $\vec{r}$ is the spatial position, $\rho(\vec{r})$ is the material's total electron density, $\rho_A^0(\vec{r})$ is the weighting factor for atom A (called the 'pro-atom density'), and $\rho^0(\vec{r})$ is the sum of pro-atom densities (called the 'pro-molecular density') for all atoms in the material.[3] For computational convenience, many stockholder charge partitioning methods define $\rho_A^0(\vec{r})$ to depend only on the distance from atom A's nucleus (i.e., $r_A$): $\rho_A^0(\vec{r}) \rightarrow \rho_A^0(r_A)$.

The Minimal Basis Iterative Stockholder (MBIS) method proposed by Verstraelen et al.[1] composes $\rho_A^0(r_A)$ as a sum of exponential functions, where the number of exponential functions equals that element's row in the Periodic Table. Specifically, one exponential function is used for H and He, two for Li through Ne, three for Na through Ar, etc. In the MBIS method, the exponential functions parameters defining $\rho_A^0(r_A)$ are optimized in a self-consistent iterative procedure.

Close examination reveals important anomalies in the article by Verstraelen et al.[1]:

(A) Verstraelen et al.[1] claimed "Because Gaussian09 does not write out the 1RDM when relativistic corrections are used, our tests on isolated molecules are limited to molecules that contain no elements heavier than krypton." (The 1RDM is a matrix representation of the electron density.) This claim is anomalous, because Gaussian09 does print an electron density matrix for DFT calculations using relativistic corrections.

(B) It is difficult for readers to assess the computational cost of the MBIS method, because the article[1] does not give information about the number of charge cycles required for convergence.

(C) In some specific instances, Verstraelen et al.[1] use non-standard terminology that creates contradictions of meaning.

(D) The reference citations in Verstraelen et al.'s article[1] are anomalous. This includes numerous key omissions.

(E) The Pareto analysis of Verstraelen et al.[1] is highly insightful and useful, but did not mention the third plot in a three variable comparison.

Each of these issues is described more fully in Section II below. The purpose of this comment article is to bring these important issues to readers' attention and to start a discussion of them.

## II. Detailed Comments

(A) On page 3898, Verstraelen et al. [1] claimed "Because Gaussian09 does not write out the 1RDM when relativistic corrections are used, our tests on isolated molecules are limited to molecules that contain no elements heavier than krypton." (The 1RDM is a matrix representation of the electron density.) This claim is anomalous, because Gaussian09 does print an electron density matrix for such DFT calculations. Verstraelen et al. [1] do not provide any supporting evidence for their claim that Gaussian09 does not write electron density matrix when relativistic corrections are used: no reference is given, no worked examples are provided, and no explanation is provided.

There are several aspects regarding this issue that need to be discussed:

(i) What is the meaning of electron density matrix when relativistic corrections are included in a quantum chemistry calculation? The Dirac equation is a fully relativistic wave equation describing electron motion.[4, 5] Because the Dirac equation involves a four-component spinor, it is cumbersome to solve for large multi-electronic systems. The Schrodinger equation is easier to solve, because it is a scalar equation, but it does not account for relativistic effects.[6] To achieve a suitable compromise between accuracy and computational burden, several schemes have been developed that add some of the terms from the Dirac equation into the Schrodinger equation as relativistic corrections. Gaussian09 integral keywords DKH, DKH0, DKHSO, and RESC can be used to include relativistic corrections.[7] In general, there are four independent components to the electron and spin magnetization density, which correspond to the four degrees of freedom of the Dirac spinor. Whenever one is dealing with an electron density matrix obtained via the Schrodinger equation or a modified Schrodinger equation to which some of the relativistic corrections have been incorporated, the electron density matrix obtained is some level of approximation to that which would be obtained through solving the fully relativistic Dirac equation.

(ii) For what kinds of calculations does Gaussian09 compute and write an electron density matrix? For non-relativistic calculations, Gaussian09 computes and writes an electron density matrix for a wide range of quantum chemistry methods including DFT, coupled cluster, configuration interaction, Hartree-Fock, MP perturbation methods, etc. For calculations including relativistic corrections, whether Gaussian09 computes the electron density matrix depends on the type of calculation requested. Since the electron density is required to compute the Coulombic repulsion between electrons, which forms an important component of the system's energy, quantum chemistry calculations that cannot compute an electron density matrix also cannot compute an energy. Such jobs terminate immediately with an error message. (I know this through personal knowledge of submitting such jobs and examining the output files.) Gaussian09 can combine implicit solvation modeling (Gaussian keyword SCRF = PCM) with relativistic corrections with DFT and Hartree-Fock methods, even using the fourth-order Douglas-Kroll-Hess method with spin-orbit coupling (Gaussian keyword Integral = DKHSO). (Relativistic calculations can also be performed without implicit solvation modeling.) Manz et al. have performed thousands of such calculations, including examples for all chemical elements atomic numbers 1 to 109 (e.g., for the computation of reference ions).[8] The Gaussian website says DKHSO is currently limited to DFT and HF methods (http://www.gaussian.com/g_tech/g_ur/k_integral.htm accessed December 31, 2016). (In agreement with the information posted on the Gaussian website, I received a termination error when attempting to perform coupled-cluster calculations combined with DKHSO in Gaussian09.)

(iii) What is the state of the art stockholder charge partitioning method for modeling diverse chemical elements? The DDEC6 method is the state of the art stockholder charge partitioning method for treating diverse chemical elements, including both light and heavy elements.[9, 10] The DDEC6 method is available for all chemical elements atomic number 1 to 109 and has been extensively tested on a wide range of main group, transition metal, lanthanide, and actinide elements. It has been successfully applied to model electron distributions of materials containing heavy elements computed with various quantum chemistry software packages, including Gaussian09[7], VASP[11, 12], etc.[9, 10] It automatically handles electron density distributions in which the core electrons were modeled using either (a) fully relaxed core electrons, (b) frozen-core electrons, or (c) effective core potentials.[9, 10, 13, 14] For calculations using (c) effective core potentials, the core electrons replaced by the effective core potential are automatically added back in at the start of charge partitioning to provide an effective all-electron charge partitioning in all cases.[9, 10, 13, 14]

Let's consider the $[GdI]^{+2}$ ion as an example to illustrate the feasibility of performing DDEC6 analysis on heavy elements using (a) fully relaxed core electrons, (b) frozen-core electrons, or (c) effective core potentials. Gadolinium has atomic number 64, and iodine has atomic number 53. All calculations performed using the PBE[15] exchange-correlation functional.

(a) fully relaxed core electrons: Using the universal Gaussian basis set of Manz and Sholl[13] combined with DKHSO, calculation first performed on Dec. 7, 2010 in Gaussian09.A.02 and repeated on Mar. 13, 2013 in Gaussian09.C.01 to generate a .wfx file. Gaussian route line as follows:

# opt PBEPBE/gen geom=connectivity scf=(fermi,maxcycle=400) NoSymm
# int=(DKHSO,NoBasisTransform,Grid=Ultrafine) output=wfx

Result gives optimized bond length of 2.71 angstroms.

(b) frozen-core electrons: Using projector augmented wave (PAW) method[16, 17], geometry optimized on Nov. 19, 2010 and electron density generated on Dec. 7, 2010 using VASP 4.6.34 and a 400 eV planewave cutoff energy. VASP includes high-level relativistic corrections for the frozen core electrons (when generating the PAW potential) and a scalar relativistic correction for the valence electrons.[18] Calculation first reported in reference [14]. Result gives optimized bond length of 2.68 angstroms.

(c) effective core potentials: Using the SDD basis set including relativistic effective core potential, calculation performed on May 31, 2011 using Gaussian09.A.02. The relativistic effective core potential replaced 28 core electrons on Gd and 46 core electrons on I. The valence electrons were treated non-relativistically. Calculation first reported in reference [14]. Result gives optimized bond length of 2.73 angstroms.

Recently, I performed DDEC6 analysis by using the Chargemol[19] code to analyze the electron densities produced by the above calculations. DDEC6 net atomic charges (NAC) and atomic spin moments (ASM) are summarized in Table 1. Note that the NACs sum to +2 and the ASMs sum to +7, which represent the net charge and overall spin magnetic moment of the molecule. As we can see, it is possible to produce and analyze an electron density matrix for heavy elements including relativistic corrections.

**Table 1**: Computed DDEC6 NACs and ASMs for $[GdI]^{+2}$

| treatment of core electrons | NACs | | ASMs | |
|---|---|---|---|---|
| | Gd | I | Gd | I |
| (a) fully relaxed | 1.98 | 0.02 | 6.84 | 0.16 |
| (b) frozen | 2.03 | -0.03 | 6.98 | 0.02 |
| (c) effective potential | 1.88 | 0.12 | 6.62 | 0.38 |

(iv) Is the MBIS method proposed by Verstraelen et al.[1] inconvenient for modeling heavy elements? Verstraelen et al.[1] claimed "Because Gaussian09 does not write out the 1RDM when relativistic corrections are used, our tests on isolated molecules are limited to molecules that contain no elements heavier than krypton." However, Gaussian09 does print an electron density matrix (aka '1RDM') for such DFT calculations. Moreover, Verstraelen et al.[1] also did not provide any examples of MBIS applied to elements heavier than krypton for the GPAW program they used in their paper to model some materials with lighter elements, so their claim of the issue being due to Gaussian09 does not stand. (In a different paper, Rogge et al.[20] used GPAW to simulate materials containing Zr (atomic number 40) and analyzed them with the MBIS method.) Moreover, as stated in (iii) above, the DDEC6 method has been successfully applied to many elements heavier than krypton, including relativistic corrections from Gaussian09 and other software packages. All of these facts suggest there is something else going on.

A close examination reveals a weakness of the MBIS method for modeling heavy elements. I am not saying it is impossible to apply the MBIS method to heavy elements, but its form makes such analysis less convenient. Consider the element uranium (atomic number 92) as an example. Since uranium is in the seventh period, the MBIS constructed pro-atom will equal the sum of seven exponential functions. Since there is one exponential function for each period, the MBIS scheme initializes these exponential functions to approximate the electrons from these respective periods.[1] Therefore, the first several exponential functions will be initialized to describe core electrons while the final one(s) will be initialized to describe valence electrons. The core electrons occupy a small spatial region around each atomic nucleus, while the valence electrons are spread over a larger volume. Moreover, core electrons should be assigned to the respective host atom, while valence electrons are partitioned among the atoms. For very heavy atoms, a large part of the MBIS fitting parameters depend on core electrons that belong to their host atoms while a smaller number of fitting parameters are devoted to the valence region that primarily governs the partitioning of electron density among atoms. Because the MBIS method does not necessarily always converge uniquely,[1] this means a change in how the core electrons are modeled during the quantum chemistry calculation that produced $\rho(\vec{r})$ could lead to significantly different computed MBIS net atomic charges. Such high sensitivity to the core electrons is not optimal.

(B) It is difficult for readers to assess the computational cost of the MBIS method, because the article [1] does not give information about the number of charge cycles required for convergence. Computational cost is one of many important criteria for judging the value of a charge partitioning method. Computational timings depend on the integration grid employed, whether the software program used a compiled or interpreted programming language, whether the computations were performed on a single computing core or in parallel across many computing cores, the processor speed, the number of charge cycles required for convergence, etc. To get a sense of the relative costs of different stockholder charge partitioning methods, it is often easiest to compare the number of charge cycles required for convergence. This makes it possible to compare results irrespective of the integration grids, serial or parallel processing, programming language and hardware architecture, because these factors should have negligible effect on the required number of charge cycles for convergence. Therefore, it is typical for articles reporting new stockholder charge partitioning methods to give several examples of the number of charge cycles required for convergence. With respect to the MBIS method, readers will want to know if more charge cycles are required for certain types of materials. For example, do MBIS computations involving an element such as potassium (atomic number 19) typically take more charge cycles to converge than computations involving hydrogen (atomic number 1), since potassium has four MBIS exponential functions and hydrogen has only one? Do computations for dense solids containing many buried atoms in the unit cell take more charge cycles to converge than small molecules that contain only surface atoms? Are there any outlying situations for which convergence is unusually slow?

(C) In some specific instances, Verstraelen et al.[1] use non-standard terminology that creates contradictions of meaning.

(i) Verstraelen et al. [1] refer to the Hirshfeld (H), Charge Model 5 (CM5), Iterative Hirshfeld (HI), DDEC4, Hirshfeld Extended (HE), Iterated Stockholder Atoms (IS)[21], and MBIS methods as 'Hirshfeld variants'. This leads to a contradictory meaning, because the Hirshfeld method is a particular type of stockholder method which has a different mathematical form than the HI, DDEC4, HE, IS, and MBIS stockholder methods. For example, the Hirshfeld method converges uniquely, while some of these other stockholder methods do not.[9] The CM5 method is not a stockholder charge partitioning method, but rather falls into category (d) mentioned in the Introduction above that assigns only an integrated net atomic charge (but not an electron density distribution) to each atom in a material. Also, the Hirshfeld and CM5 methods require only a single isolated reference atom density for each atom in a material, while the HI, DDEC4, and HE methods require a reference ion library of various oxidation states for each chemical element, and the IS and MBIS methods do not require any quantum mechanically computed reference ions. To avoid confusion, it is best to call stockholder charge partitioning methods 'stockholder charge partitioning methods'. The term 'Hirshfeld variants' should be reserved for different implementations of the Hirshfeld method, such as those which use slightly different schemes for computing the neutral reference atoms. For example, one Hirshfeld variant may compute the neutral reference atom in the complete basis set limit using a single chosen exchange-correlation theory irrespective of the level of theory used to compute the material's electron distribution, while another Hirshfeld variant may use the same basis set and exchange-correlation theory to compute the neutral reference atoms and the material's electron distribution. Analogously, there are several different Iterative Hirshfeld variants that use different schemes to compute the reference ion library.[22-24] Referring to other stockholder charge partitioning methods as 'Hirshfeld variants' will produce massive confusion, because mathematical analysis results that apply to the Hirshfeld method will not necessarily apply to these other methods which are described by different functional forms.

(ii) In Section 5, Verstraelen et al.[1] incorrectly use the term "spherical closed shell (SCS) atoms" to denote atoms that may contain partially filled valence shells. In quantum chemistry, the term 'closed shell' denotes a filled shell, while the term 'open shell' denotes a partially filled shell. The origination of this terminology is that a filled shell is 'closed' because it cannot accept any more electrons, while a partially filled shell is 'open' because it can accept more electrons. Examples of closed shell atoms include the isolated neutral noble gas atoms He, Ne, Ar, Kr, etc. Examples of open shell atoms include the isolated neutral alkali metal atoms Li, Na, K, etc. What Verstraelen et al.[1] refer to as "spherical closed shell atoms" is properly called 'spin unpolarized calculation of atoms constrained to spherical symmetry'. Specifically, these spin unpolarized calculations had fractionally occupied valence orbitals when needed to create spherically symmetric atoms with no spin polarization.[1]

(D) The reference citations in Verstraelen et al.'s article[1] are anomalous. This includes numerous key omissions.

(i) On page 3895, Verstraelen et al. state "Unfortunately, the HI method also has its deficiencies." However, they do not mention the spontaneous symmetry breaking problem in which the HI method sometimes assigns vastly different net atomic charge values to symmetry equivalent atoms.[9] This spontaneous symmetry breaking is due to the HI method having a non-convex optimization landscape that produces non-unique solutions.[9] This problem is certainly one of the major flaws of the HI method and deserved mentioning.

(ii) Their article[1] does not cite the most state-of-the-art stockholder charge partitioning method available. While the authors were performing their work, the DDEC6 method was published, which yields excellent results for a wider range of material types than all other stockholder charge partitioning methods developed to date.[9, 10] Some of the authors of reference[1] had the benefit of reading the DDEC6 manuscripts and articles before their article was submitted for publication. Even if their development of the MBIS method was well underway at this time, they still had the benefit of seeing the presentation of results and wording of important issues in the DDEC6 manuscripts and articles, which they could use to refine the presentation of their MBIS manuscript. Moreover, it is reasonable for readers to expect them to cite the current state-of-the-art. Their omission of citations to the DDEC6 articles is noteworthy.

(iii) There are anomalous aspects to the authors' claims in the following paragraph:

"The basis set robustness of the H, HI, and HE methods can be improved as follows. Currently, we have used consistent levels of theory for pro-atom and molecular electron densities. If the pro-atoms were computed with a single level of theory and basis set, independent of the settings of the molecular calculation, the robustness would significantly improve. This is noticeable in the low sensitivity of the DDEC4 and especially the CM5 charges. Both CM5 and DDEC are implemented with a unique set of pro-atoms." (reference [1] pages 3905-3906)

Compare the above paragraph to the following passage in an arXiv preprint by Manz and Gabaldon-Limas:

> "This definition is motivated by practical considerations. First, requiring the reference ions to be converged using the same basis set family as employed in the quantum mechanical calculation of the system's total electron density distribution is not optimal. For example, an atom-centered basis set for a large cluster of water molecules (e.g., aug-cc-pvtz) is more complete than the same family of atom-centered basis set (e.g., aug-cc-pvtz) applied to a single isolated atom. Consequently, requiring the aug-cc-pvtz basis set to be used for each isolated reference ion would be requiring the reference ions to be converged using a less complete basis set than that used for the polyatomic system. Therefore, we believe the most appropriate approach is to compute the reference ions near the complete basis set limit.
>
> Second, it is impractical to require the reference ions to be converged using the same exchange correlation (XC) theory as used in the quantum mechanical calculation of the system's total electron density distribution. Converging the reference ions is a tricky and time-consuming process that is best completed off-line rather than on-the-fly at the beginning of a charge partitioning calculation. Aside from the nearly impossible task of optimizing the reference ions separately for each of the countless different XC theories, there are theoretical motivations for using a single fixed reference ion library. Consider a situation in which the same material M is studied with two different XC theories called XC1 and XC2. Suppose we construct some quantitative measure of similarity between two electron distributions and use it to quantify (a) how similar the electron distribution of the material M computed with XC1 is to the electron distribution of the material M computed with XC2 and (b) how similar a reference ion library (containing the chemical elements in M) computed with XC1 is to a reference ion library computed with XC2. There are three possibilities: (i) XC1 and XC2 give much more similar electron distributions for the reference ions than they do for the material M, (ii) XC1 and XC2 give much more similar electron distributions for the material M than they do for the reference ions, and (iii) XC1 and XC2 give electron distributions of comparable similarity for the material M and the reference ions. In case (i), XC1 and XC2 give similar reference ion densities compared to their different electron distributions for the material M, so in this case we can confidently use the XC1 reference ions to analyze the electron distribution of material M computed via either XC1 or XC2. In case (ii), XC1 and XC2 give similar electron distributions for the material M compared to their different reference ion densities, so in this case the use of two different reference ion sets (i.e., XC1 and XC2) would introduce an artificial difference in NACs for the material that reflects more the change in reference ion sets than the change in the material's electron distribution. Accordingly, in case (ii), we would be better off to choose one of the reference ion sets (e.g., XC1) and use it consistently to analyze the material's electron distribution computed via either XC1 or XC2. In case (iii), where the changes in reference ion densities between XC1 and XC2 are of comparable magnitude to the changes in the material's electron density, it cannot be determined a priori whether using a fixed reference ion library (i.e., using XC1 reference ions to analyze the material's electron distribution computed with either XC1 or XC2) or a variable reference ion library (i.e., using the same XC functional to compute the reference ions as was used to compute the material's electron distribution) is better. While cases (i) and (iii) are not decisive, case (ii) clearly favors using a fixed reference ion library computed with one XC functional irrespective of the XC functional used to compute the material's electron distribution. Therefore, we believe the most appropriate approach is to define the reference ions using a specific XC theory (i.e., PW91) irrespective of the XC theory employed in the quantum mechanical calculation of the system's total electron density distribution." (reference [8] pages 41-42)

As far as I am aware, the arXiv preprint quoted above is the first time that anyone has proposed that the reference ions be computed using a single exchange-correlation theory and basis set to improve the transferability of the computed stockholder results. Previous reference ion libraries using a single exchange-correlation theory and basis set were motivated primarily by simplicity considerations (i.e., it is easier to compute and store the reference ion library one time rather than recompute it with different basis sets and exchange-correlation theories), but not by the intentional design to improve the transferability of the computed stockholder results.[13] The idea to compute the reference ion library using a single exchange-correlation theory and basis set (i.e., near the complete basis set limit) to improve the transferability of the computed stockholder results was originated by Tom Manz.[8] Moreover, the first complete such reference ion library (for elements 1 to 109) was computed by Manz and Limas[8] using the method and some of the results from Manz and Sholl[13, 14].

The originators of the CM5[25] method envisioned recomputing the neutral reference atoms using the same exchange-correlation theory and basis set as employed in the quantum chemistry calculation of the material's electron distribution, as explicitly stated and explained on page 5643 of Wang et al.[26] (Like the Hirshfeld method, the CM5 method uses only neutral reference atoms and does not employ any charged reference ions.[25]) Specifically, Wang et al.[26] stated that either fixed or recomputed reference atoms could be used to compute the CM5 charges, but they explicitly favored recomputing the reference atoms using the same exchange-correlation theory and basis set as employed in the quantum chemistry calculation of the material's electron distribution.

Verstraelen et al.[1] do not cite the Manz and Limas arXiv preprint[8], even though some of the authors of reference [1] had access to it (or a previous manuscript containing the above quoted passage) before their article was submitted for publication.

(iv) On page 3907, the authors claim, without presenting sufficient supporting evidence, that the use of the "spherical closed shell (SCS)" reference atoms is "the most common in the context of of dispersion models." As I have pointed out above, what Verstraelen et al. called "spherical closed shell (SCS) atoms" is a misnomer that actually means 'spin unpolarized calculation of atoms constrained to spherical symmetry'. Besides the incorrect terminology, there is the additional problem that Verstraelen et al.[1] do not back their claim with sufficient supporting evidence. Verstraelen et al.[1] claim Tkatchenko and Scheffler[27] used spin unpolarized reference atoms, but that paper[27] does not indicate whether the reference atoms were obtained via spin polarized or unpolarized calculations. If Verstraelen et al.[1] had knowledge of how those reference atoms were computed (for example, through correspondence with Tkatchenko and Scheffler, another reference by Tkatchenko and Scheffler, or by examination of Tkatchenko and Scheffler's software codes), they should have indicated this. Moreover, other authors besides Tkatchenko and Scheffler have applied dispersion corrections and some of these used spin polarized calculations for the reference atoms.[28] (I know through correspondence with authors of reference [28] that they employed spin-polarized calculations of the reference atoms to compute the free atom volumes.) Because many papers on dispersion corrections do not explicitly state whether the reference atoms were computed using spin polarized or unpolarized calculations, it is difficult to make a judgement as to which is the most common.

(v) Excessive self-citing and excessive minimization of the contributions of competitors. For example, on page 3895 Verstraelen et al.[1] state "Unfortunately, also the HI method has its deficiencies. For example, when the method is applied to highly polar oxides, it requires the spherically averaged density of the nonexisting oxygen dianion as input.[29]" The manner in which this citation is constructed makes it appear that this is something that was identified by Verstraelen et al. in one of their prior works [29]. In fact, this was not originated by Verstraelen et al. but rather by Manz and Sholl[13, 14]. Of crucial importance, it was Manz and Sholl who first introduced charge compensated reference atomic ions into stockholder charge partitioning methods including iterative Hirshfeld and Density Derived Electrostatic and Chemical (DDEC) methods to describe charged atoms in materials, especially ions like the oxygen -2 anion in order to stabilize their charges.[13] The specific case of the oxygen -2 anion was mentioned in the work of Manz and Sholl[13, 14], as it had also been in a much earlier work by Watson that pre-dates stockholder charge partitioning methods[30]. Manz and Sholl introduced two distinct methods of charge compensation to describe electrostatic screening and dielectric effects.[13] The work of Verstraelen et al.[29] which appeared several years later did not first identify the problem stated above.

When citing their competitors, Verstraelen et al.[1] often state methods developed by their competitors are "variants" of someone else's work. For example, on page 3902 they state "A particular improvement of HI over Hirshfeld is that HI charges make a good estimate of the electrostatic potential of organic molecules.[31] This is no longer the case for metal oxides, e.g., the ESP in the pores of zeolites, which inspired several groups to further improve the method, leading to variants such as Hirshfeld-E[29] (HE) and DDEC4.[14]" The problem with such wording is that it fails to recognize the important independent contributions and the numerous crucial insights that led to developing new methodologies. The main importance of the DDEC methodologies is not that they are "variants" of other people's work, but rather that they have numerous key important and crucial new insights and computational advances that allow for stockholder atomic population analysis to be applied more accurately to a wider range of material types.[9, 10, 13, 14, 28, 32, 33] The DDEC methodologies do not consist of one improvement which was inspired by one insight over previous methods, rather they contain numerous crucial new key insights, with subsequent generations incorporating more key insights that lead to better performance across a wider range of material types.[9, 10]

The above are representative examples, and I am not trying here to compose an exhaustive list, which would be tedious.

(E) The Pareto analysis of Verstraelen et al.[1] is highly insightful and useful, but did not mention the third plot in a three variable comparison. In Section 4.6 of their article[1], Verstraelen et al. use Pareto analysis to compare tradeoffs between the root mean squared error (RMSE) of electrostatic potential and interaction energy (Fig. 9a of reference[1]) and between sensitivity of the net atomic charges and RMSE of interaction energy (Fig. 9b of reference[1]). (Here, the sensitivity of the NACs is a type of inverse of the transferability of the NACs.[1]) The Pareto front for the third tradeoff between these variables (i.e., between RMSE of electrostatic potential and sensitivity of the NACs) was not discussed. However, I reconstructed it from the data listed in Figs. 5 and 8 of reference[1]. As shown in Figure 1 below, this Pareto front contains the CM5, MBIS, DDEC4, IS, CHELPG, MSK, and HLY methods in this order from lowest sensitivity to lowest RMSE of electrostatic potential. The Lowdin, Mulliken (off scale), and QTAIM (off scale) methods performed extremely poorly. Interestingly, the RESP method, which was introduced to make the electrostatic potential fitting charges more transferable,[34] actually did not achieve a favorable combination of charge transferability and accuracy for fitting the electrostatic potential.

This Pareto plot, together with Figs. 9a and 9b of reference[1] reveals a fundamental flaw in a previous paper by Verstraelen et al.[29] Specifically, the paper entitled "Hirshfeld-E Partitioning: AIM Charges with an Improved Trade-Off between Robustness and Accurate Electrostatics" introduced a new stockholder charge partitioning method (called Hirshfeld-E and abbreviated HE) that Verstraelen et al.[29] claimed produced an improved trade-off between transferability ("robustness") of the NACs and the accuracy of describing the electrostatic potential surrounding the material. In the conclusion of that paper, Verstraelen et al.[29] stated "Therefore, the HE scheme is an attractive new method for the development of accurate and transferable electrostatic force-field terms, especially for polar systems." The Pareto plots show this conclusion to be fundamentally wrong. As shown in Figure 1 below, the HE scheme has remarkably worse transferability of NACs compared to the HI scheme, while providing about the same overall accuracy for reproducing the electrostatic potential surrounding materials. As shown in Figs. 9a and 9b of reference[1], the HE method also performs significantly worse than the HI method for the RMSE of electrostatic interaction energy in molecular dimers.

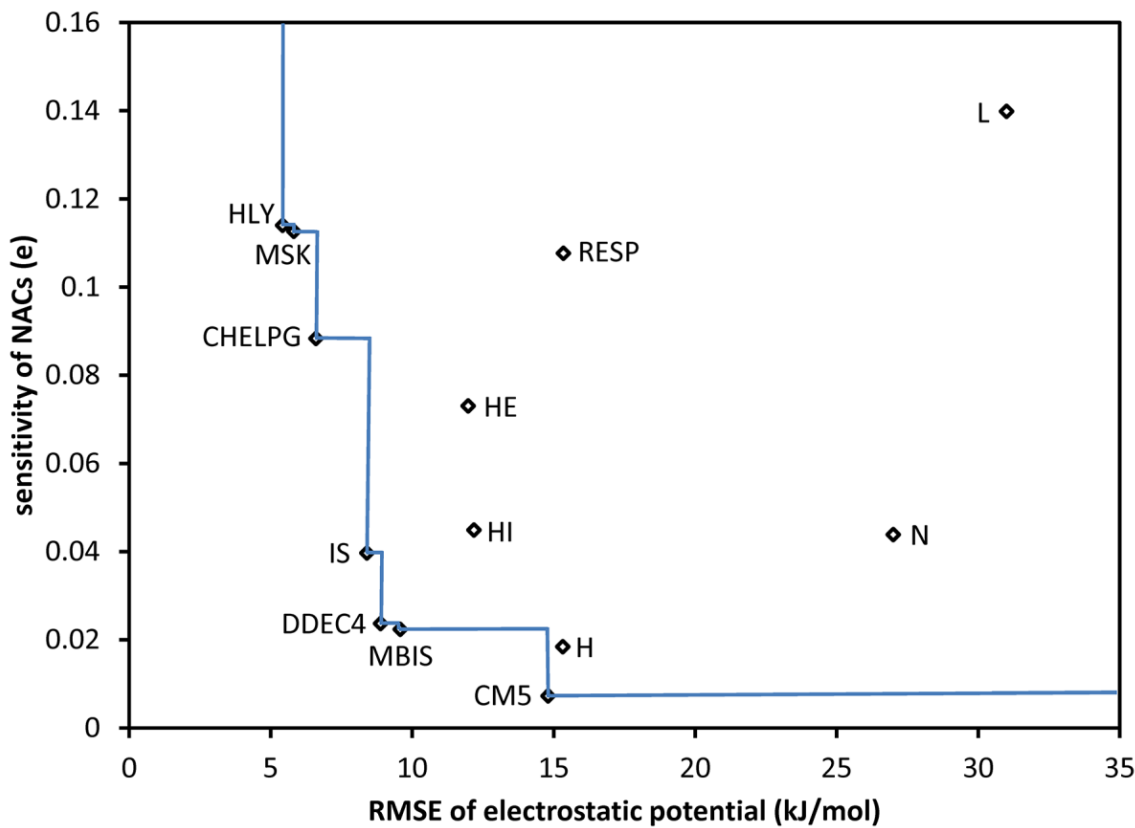

Figure 1: Pareto plot for sensitivity of NACs versus RMSE of electrostatic potential. The Mulliken and QTAIM methods performed so poorly that they are off the scale of the plot.

Verstraelen et al.[29] reached an incorrect conclusion in their previous work introducing the Hirshfeld-E method, because the test set they employed was not chemically diverse. Specifically, when introducing the Hirshfeld-E method, Verstraelen et al.[29] used only a test set containing 248 hydrogen terminated silica clusters. This test set contained silica clusters with two to eight silicon atoms, in which each silicon atom was bound to four other atoms (either O or H), each oxygen atom was bound between two silicon atoms, the cluster was terminated by hydrogen atoms, rings in the bond graph with less than four silicon atoms were not allowed, and the optimized Si-O bond length deviated by no more than 10% from the value for silica.[29] In such clusters, the atoms have oxidation state of +4 (Si), -2 (O), and -1 (H). It was not the number of materials in the test set, but rather the chemical similarity of those materials, that led Verstraelen et al.[29] to the incorrect conclusion. Specifically, their test set included only three chemical elements and each of those chemical elements was present in only a single oxidation state within the test set's materials.[29] The MBIS article by Verstraelen et al.[1] includes multiple material test sets and shows the Hirshfeld-E method does not offer an improved combination of transferability and accuracy for reproducing electrostatics. Notably, these test sets included the hydrogen terminated silica clusters along with numerous additional material test sets representing organic molecules and inorganic clusters.[1]

The fundamental lesson is that it is crucial to include a chemically diverse range of materials when introducing new charge assignment methods. I believe the Hirshfeld-E paper[29] should not have passed peer review, because it included only a single chemically non-diverse material test set. The MBIS paper is much better in this regard, because it includes numerous different test sets spanning a more diverse range of materials.[1] However, there are insufficiently addressed questions about how MBIS performs for heavy elements and dense solids. The MBIS method was tested for a few porous and nonporous solids,[1] but the number was too few to draw general conclusions about how MBIS performs for dense solids.

### III. Good Points

Verstraelen et al.'s article [1] has several good points. One of the strongest points of the article is its use of various datasets to compare fourteen different charge assignment methods. This is commendable in three ways. First, the charge assignment methods included key representatives from each of the four major classes defined in the Introduction above. The second key point is that they used several different datasets containing numerous organic molecules and inorganic clusters to statistically quantify the performance of these different charge assignment methods. The third point is that they analyzed the performance of these methods for (a) how accurately they reproduced the electrostatic potential surrounding the molecule or cluster, (b) how accurately they reproduced the intermolecular electrostatic interaction energy between a pair of molecules (including datasets with hydrogen bonds, halogen bonds, and other diverse non-covalent interactions), and (c) the transferability of the computed net atomic charges. The figures in the paper were particularly well done and presented the data in ways that are insightful and easy to digest.

Another strong point of the article is its consideration of the spherical charge penetration effect. What is particularly interesting about this aspect is they have shown in Figure 9 of reference [1] that the spherical charge penetration effect is most noticeable in the RMSE of the interaction energy between molecules in the molecular dimers while not as noticeable in the RMSE of the electrostatic potential surrounding a molecule. This is interesting to me, because of my previous work which found spherical charge penetration had little effect on the RMSE of the electrostatic potential.[9, 10] Knowing that the spherical charge penetration effect is more important for the electrostatic interaction within a molecular dimer will be helpful to my future research.

Figure 9a of reference [1] raises some interesting issues regarding the distinction between RMSE in the electrostatic potential surrounding a molecule and the electrostatic interaction energy between molecules in a dimer. Of particular interest is that natural population analysis exhibited a high RMSE in the electrostatic potential but a low RMSE in the interaction energy. It would be interesting to learn more about why this is the case.

### Acknowledgement

The computational results presented in Table 1 used supercomputing resources provided by the Extreme Science and Engineering Discovery Environment (XSEDE). XSEDE is funded by NSF grant ACI-1053575. XSEDE project grant TG-CTS100027 provided resource allocations.